\documentclass[nocompress]{spie}  
\usepackage[]{xspace,amsmath,graphicx}
\usepackage[caption=false]{subfig}
\newcommand{\FIG}[3]{\includegraphics[width=#1\linewidth,draft=#2]{#3.eps}}

\newcommand{\loD}{\mbox{$\lambda/D$}\xspace}
\newcommand{\e}[1]{10^{#1}}
\newcommand{\E}[1]{\times10^{#1}}
\newcommand{\Coro}{Coronagraph\xspace}

\newcommand{\coro}{coronagraph\xspace}
\newcommand{\coros}{coronagraphs\xspace}
\newcommand{\wfc}{wavefront control\xspace}

\graphicspath{{D:/Documents/2012a_EXCEDE/DATA/}{D:/Documents/2012a_EXCEDE/REPORTS/}}

\title{Experimental study of a low-order wavefront sensor for high-contrast coronagraphic imagers: results in air and in vacuum} 

\author{Julien~Lozi\supit{a}, Ruslan~Belikov\supit{b}, Sandrine~J.~Thomas\supit{b,c}, Eugene~Pluzhnik\supit{b,c}, Eduardo Bendek\supit{b}, Olivier~Guyon\supit{a}, Glenn~Schneider\supit{a}\skiplinehalf
\supit{a}University of Arizona, 1401 E University Blvd, Tucson, AZ 85721, USA; \\
\supit{b}NASA Ames Research Center, Moffett Field, CA 94035, USA; \\
\supit{c}UARC/NASA Ames, P.O. Box 7, Moffett Field, CA 94035, USA
}

\authorinfo{Further author information: (Send correspondence to J.L.)\\ J.L.: E-mail: jlozi@email.arizona.edu, Telephone: 1 650 604 6549}

\begin{document} 
\maketitle 

\begin{abstract}
\end{abstract}

For the technology development of the mission EXCEDE (EXoplanetary Circumstellar Environments and Disk Explorer) ---~a 0.7~m telescope equipped with a Phase-Induced Amplitude Apodization \Coro (PIAA-C) and a 2000-element MEMS deformable mirror, capable of raw contrasts of $\e{-6}$ at 1.2 ~\loD and $\e{-7}$ above 2~\loD~--- we developed two test benches simulating its key components, one in air, the other in vacuum. To achieve this level of contrast, one of the main goals is to remove low-order aberrations, using a Low-Order WaveFront Sensor (LOWFS). We tested this key component, together with the \coro and the \wfc, in air at NASA Ames Research Center and in vacuum at Lockheed Martin. The LOWFS, controlling tip/tilt modes in real time at 1~kHz, allowed us to reduce the disturbances in air to $\e{-3}$~\loD rms, letting us achieve a contrast of $2.8\E{-7}$ between 1.2 and 2~\loD. Tests are currently being performed to achieve the same or a better level of correction in vacuum. With those results, and by comparing them to simulations, we are able to deduce its performances on different \coros ---~different sizes of telescopes, inner working angles, contrasts, etc.~--- and therefore study its contribution beyond EXCEDE.




\section{INTRODUCTION}
\label{sec:intro}

If we want to take a full advantage of future space \coros, we will have to achieve deeper contrast at small Inner Working Angles (IWA). But even in space, experience showed that telescopes are affected by the vibrations of the reaction wheels, as well as pointing stability. Therefore, if we want to image planets and stellar environments at 1~AU, \coros will have to be equipped with Low-Order Wavefront Sensors (LOWFS) coupled with fast steering mirrors (FSM) and deformable mirrors (DM), capable of measuring and correcting low-order aberrations.

In the context of the EXCEDE (EXoplanetary Circumstellar Environments and Disk Explorer) mission\cite{Guyon12a}, we are currently testing its starlight suppression system (the \coro with the wavefront correction and the LOWFS) in air, at the Ames Coronagraphic Experiment (ACE) at NASA Ames Research Center (Moffett Field, CA) and in vacuum at Lockheed Martin (Palo Alto, CA). A description of the challenges faced in the mechanical design of the vacuum testbench is presented during this conference\cite{Bendek14}, as well as the contrast results we obtained in air and in vacuum, and their future implication\cite{Belikov14}.

In this paper, I will recap the performances of the Low Order WaveFront Sensor and Control (LOWFSC) of our testbench in air in Sec.~\ref{sec:ReminderOfTheResultsObtainedInAir}, then I will describe the setup of the vacuum version of the LOWFSC in Sec.~\ref{sec:Setup}. Section~\ref{sec:AnalysisOfFastInstabilities} will show an analysis of the fast instabilities measured on the bench, while Sec.~\ref{sec:ShortTermDrift} and Sec.~\ref{sec:LongTermInstabilities} while describe more short term and long term drifts. 


\section{Reminder of the results obtained in air}
\label{sec:ReminderOfTheResultsObtainedInAir}

The principle and the setup of the LOWFSC used in air have already been described in~\cite{Lozi13}. During our tests in air, we demonstrated that the LOWFSC using piezo actuators could reduce tip/tilt aberration down to $1.5\E{-3}$~\loD rms at 1~kHz. The residue was limited by mechanical vibrations of optical elements, while the measurement noise was around $4\E{-4}$~\loD rms when the flux is optimal.

In these conditions, we achieved with speckle nulling a contrast level of $1.8\E{-7}$ between 1.2 and 2~\loD, limited by low order aberrations, and $6.5\E{-8}$ between 2 and 4~\loD, probably limited by incoherent light.


\section{Setup in vacuum}
\label{sec:Setup}

For our tests in vacuum, the setup is a bit different than in air. This setup is closer to the proposed EXCEDE mission: the DM is downstream of the Phase-Induced Amplitude Apodization (PIAA) mirrors, and a set of inverse PIAA lenses is used to increase the Outer Working Angle (OWA). The Focal Plane Mask (FPM) used here is not the final one: its size is 20~times too big, so we have to translate it in one axis to have an IWA of 1.2~\loD on that axis. We define the IWA as the minimum angle where the potential planet has a transmission of 50\%. For more information about the mechanical setup, see \cite{Belikov14,Bendek14}

The LOWFSC setup is very similar to the one described in~\cite{Lozi13}: a LOWFS camera ---~actually the same than our test in air, but this time prepared for vacuum~--- is used to measure the displacement of the PSF in the focal plane of the PIAA. The difference here is that the FPM is not a three-zone mask as described in~\cite{Lozi13}, so we might have unseen drifts of the FPM. Also, for this vacuum test, we didn't have any fast actuator, like the piezoelectric translations used in air, or a fast steering mirror. The control is therefore performed using the DM itself, by sending tip/tilt commands to it, at a frequency of 1~Hz.


\section{Analysis of Fast Instabilities}
\label{sec:AnalysisOfFastInstabilities}

During our experiments in air, the main disturbances observed on the tip/tilt had two origins: a residual turbulence due to small gradients of temperature inside the bench, and vibrations of the different optical elements, and the bench itself. By going in vacuum, we removed the turbulence contribution, but we still observe some vibrations. Also we discovered a very slow drift, that I will describe in Sec.~\ref{sec:LongTermInstabilities}. In this section I will describe our vibration analysis. 

For those measurements, I recorded 10000~images at 1~kHz with the LOWFS. For those measurements, the noise level is below $\e{-3}$~\loD. I measured the stability in four different cases:
\begin{enumerate}
	\item When the cryo pump and the cooling system of the cameras are running,
	\item when I turn off the cooling system,
	\item when I turn off the cryo pump,
	\item when I turn off both the cryo pump and the cooling system.
\end{enumerate}

\begin{figure}[t]%
\center
\FIG{0.49}{false}{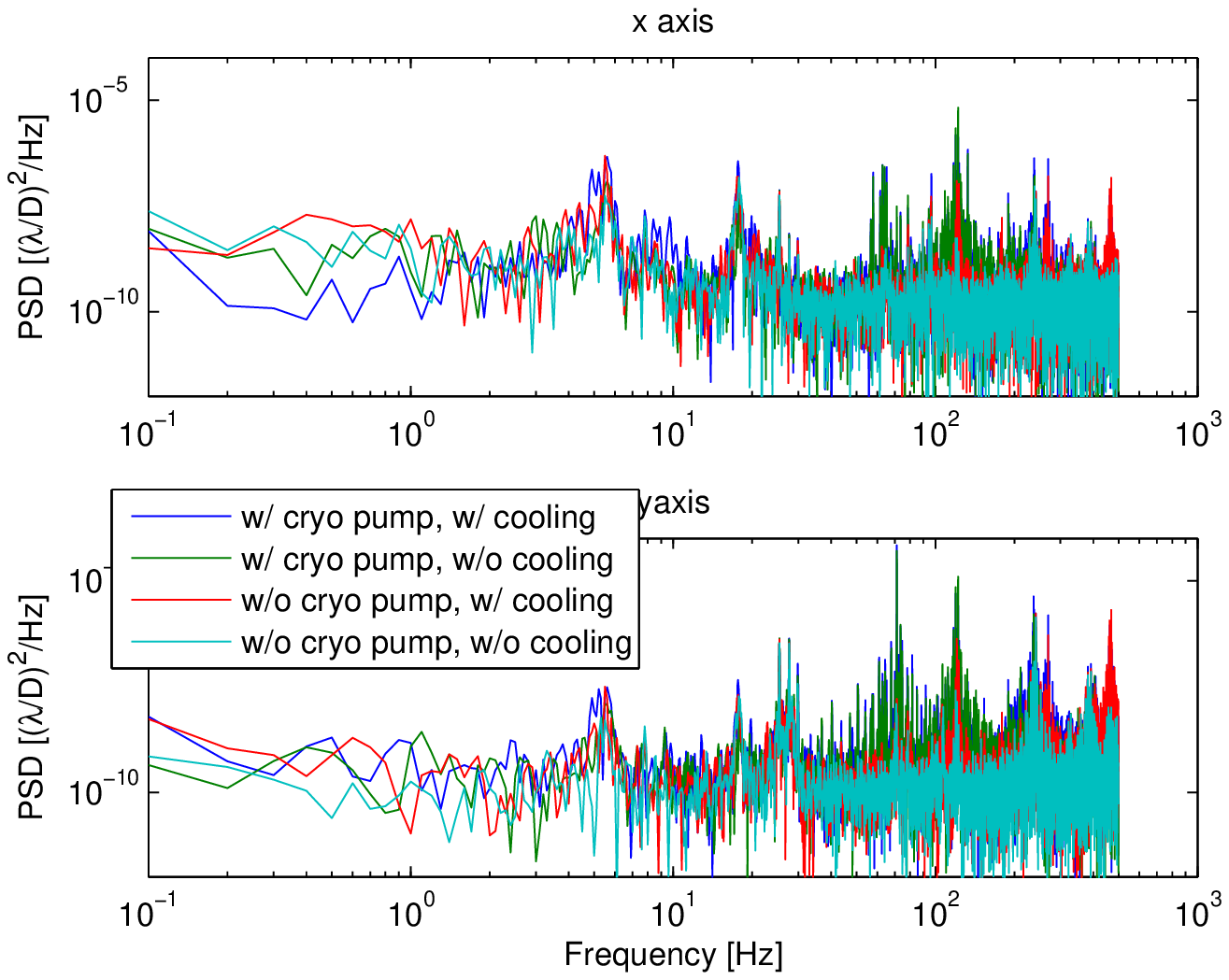}
\FIG{0.49}{false}{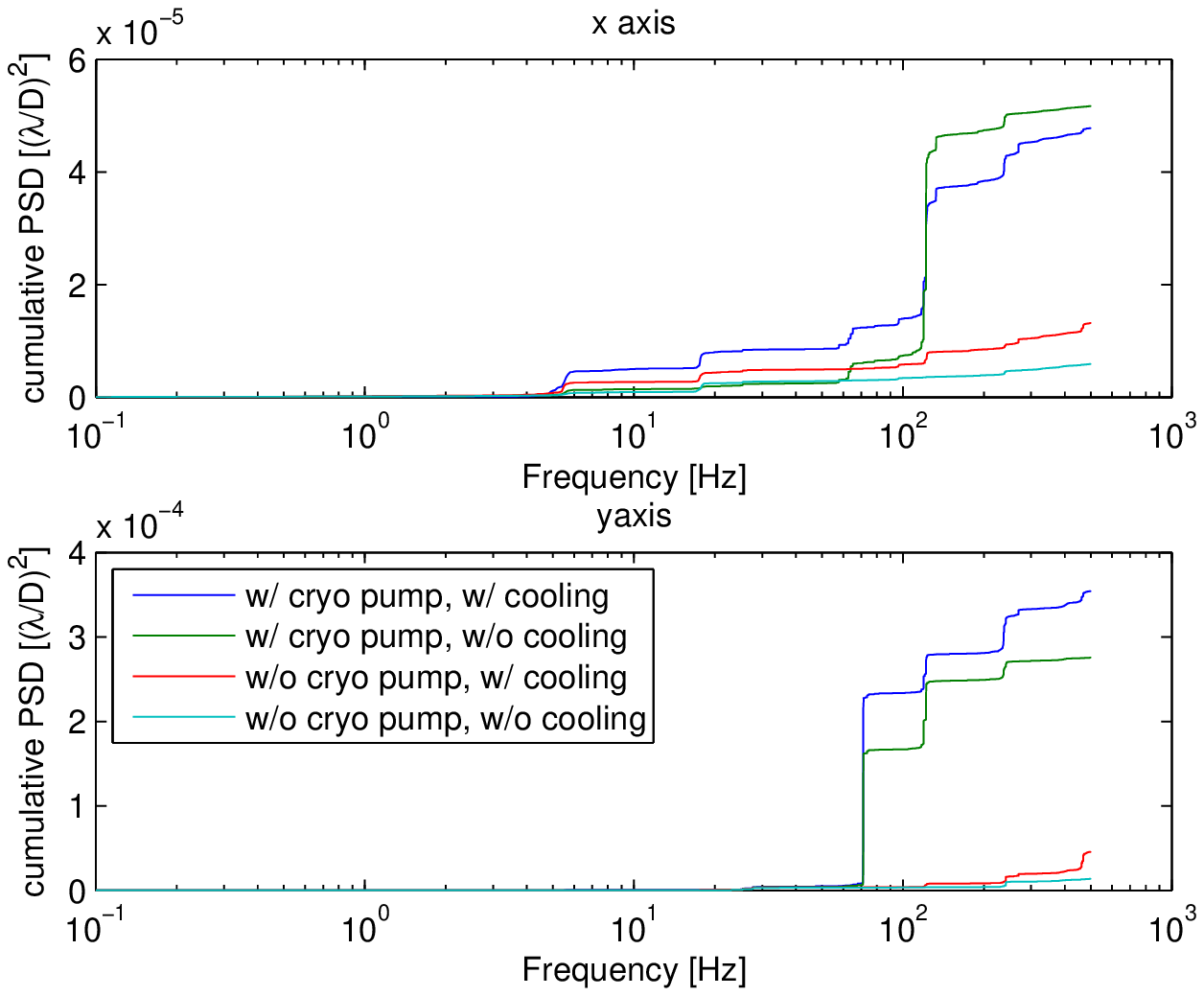}
\caption{Stability of the PSF after the PIAA mirrors in vacuum test \#3, with the cryo pump and/or the cooling system off: PSD (left) and cumulative PSD (right).}%
\label{fig:stabPIAAvacpump1}%
\end{figure}

Figure~\ref{fig:stabPIAAvacpump1} presents the Power Spectrum Densities (PSDs), and the cumulative PSDs of the measurements for the different cases. The cryo pump amplifies mostly vibrations at 122, 241, 71 and 5.5~Hz on the x-axis, and vibrations at 71, 122 and 241~Hz on the y-axis. The cooling system amplifies also a little bit the vibration at 122~Hz on both axes, as well as the 5.5~Hz in x, but it also amplifies vibrations at 269 and 469~Hz on both axes.

\begin{table}[t] \centering
  \caption{Vibration contribution over 10~s in the different cases.}
  \medskip
  \begin{tabular}{ccc}
    \hline \hline
    Case & x-axis & y-axis \\
    \hline
    \#1 & $5.8\E{-3}$~\loD & $1.58\E{-2}$~\loD \\
	  \#2 & $5.9\E{-3}$~\loD & $1.34\E{-2}$~\loD \\
	  \#3 & $3.3\E{-3}$~\loD & $5.6\E{-3}$~\loD \\
	  \#4 & $2.0\E{-3}$~\loD & $3.0\E{-3}$~\loD \\
    \hline \hline
  \end{tabular}
  \label{tab:vibrations}
\end{table}

The measured stability are presented in Tab.~\ref{tab:vibrations}. In this table, we can see that the main contributor to the vibrations is created by the cryo pump, especially on the y-axis: between cases~\#1 and~\#3, the stability is reduced by a factor~1.8 in x, and a factor~2.8 in y. If we also turn off the cooling system of our cameras, the stability is then reduced by another factor~1.7 in x, and a factor~1.9 in y. Without the cryo pump, We can maintain the vacuum level in the chamber using only our turbo-molecular pump. But we cannot perform long-term measurements without the cooling system of the cameras, because they overheat a lot without it. So our nominal case for the contrast measurements is the case~\#3. In this experiment, the vibration level is higher than our bench in air, where we obtained a closed-loop stability ---~i.e. so without the turbulence~--- of $1.5\E{-3}$~\loD in x, and $2.0\E{-3}$~\loD in y.

Some of those vibrations are probably due to a an unstable mounting of the LOWFS camera, so they might correspond to non-common path errors if we try to correct them. In our first vacuum test, the mounting solution of the LOWFS camera was a bit different, and we obtained a stability of $1.8\E{-3}$~\loD on both axes, so approx. the same level as in our testbench in air. We are currently improving the mounts of the LOWFS and the FPM to guaranty a better stability.


\section{Short-Term Drift}
\label{sec:ShortTermDrift}

During the first vacuum test, both the LOWFS and the science camera temperatures where not controlled by the cooling system. We observed a drift of the PSF in front of both cameras, during the acquisition. To determine and quantify that drift, I recorded series of images during one to two hours at low frequency: 1~Hz for the science camera, and 10~Hz for the LOWFS camera. 

For both cameras, I analyzed the drift by measuring the centroid of the PSFs. For each image, I correct the background increase of the background (due to the temperature increase), I also correct for the intensity variations, and I apply a threshold, to remove the noise of the background.

During these measurements, I also recorded the different temperatures available: the temperature of the detector and the heat sink for the science camera, and the temperature of the detector for the LOWFS camera.

\begin{figure}[t]%
\center
\FIG{0.49}{false}{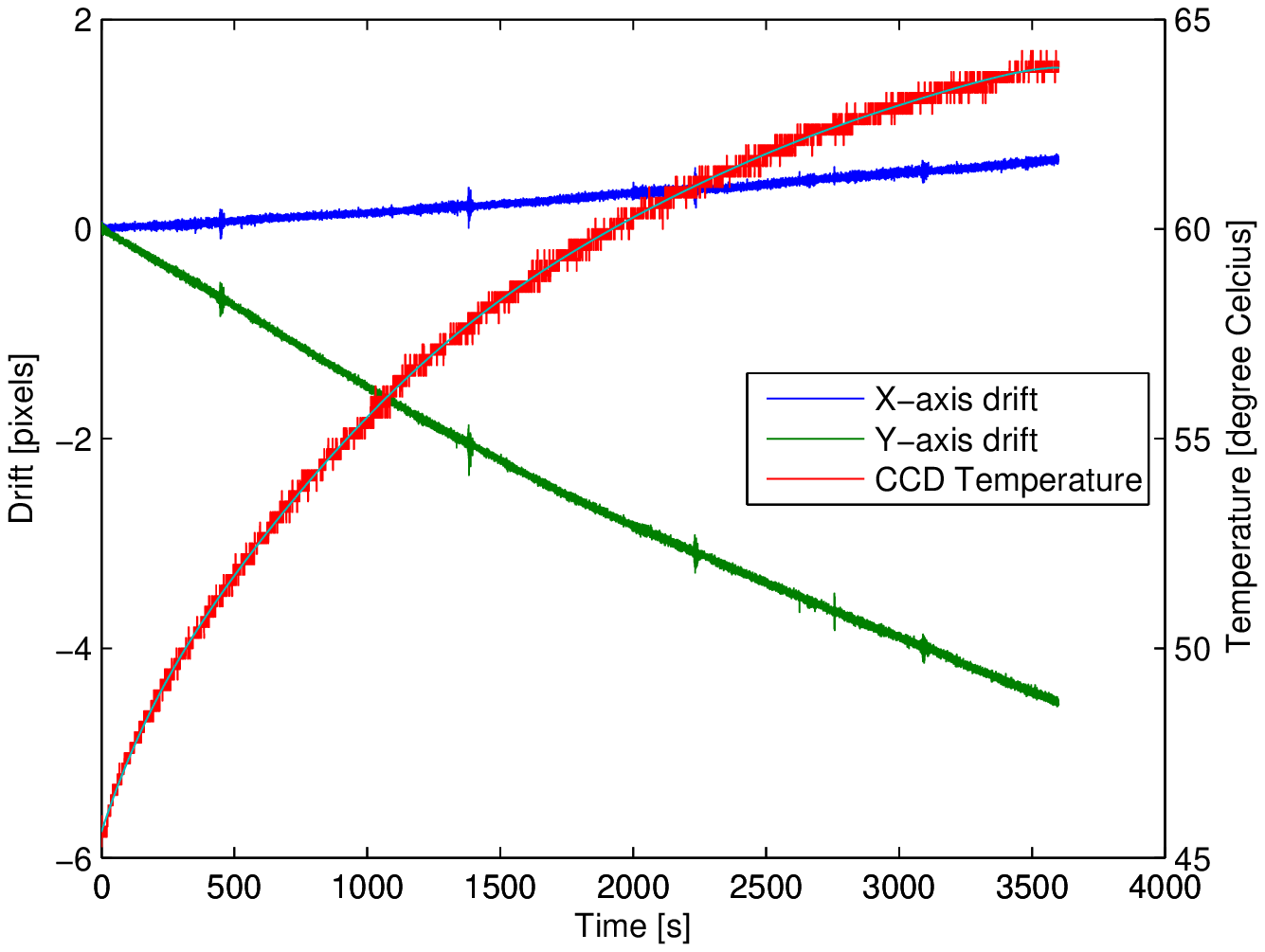}
\FIG{0.49}{false}{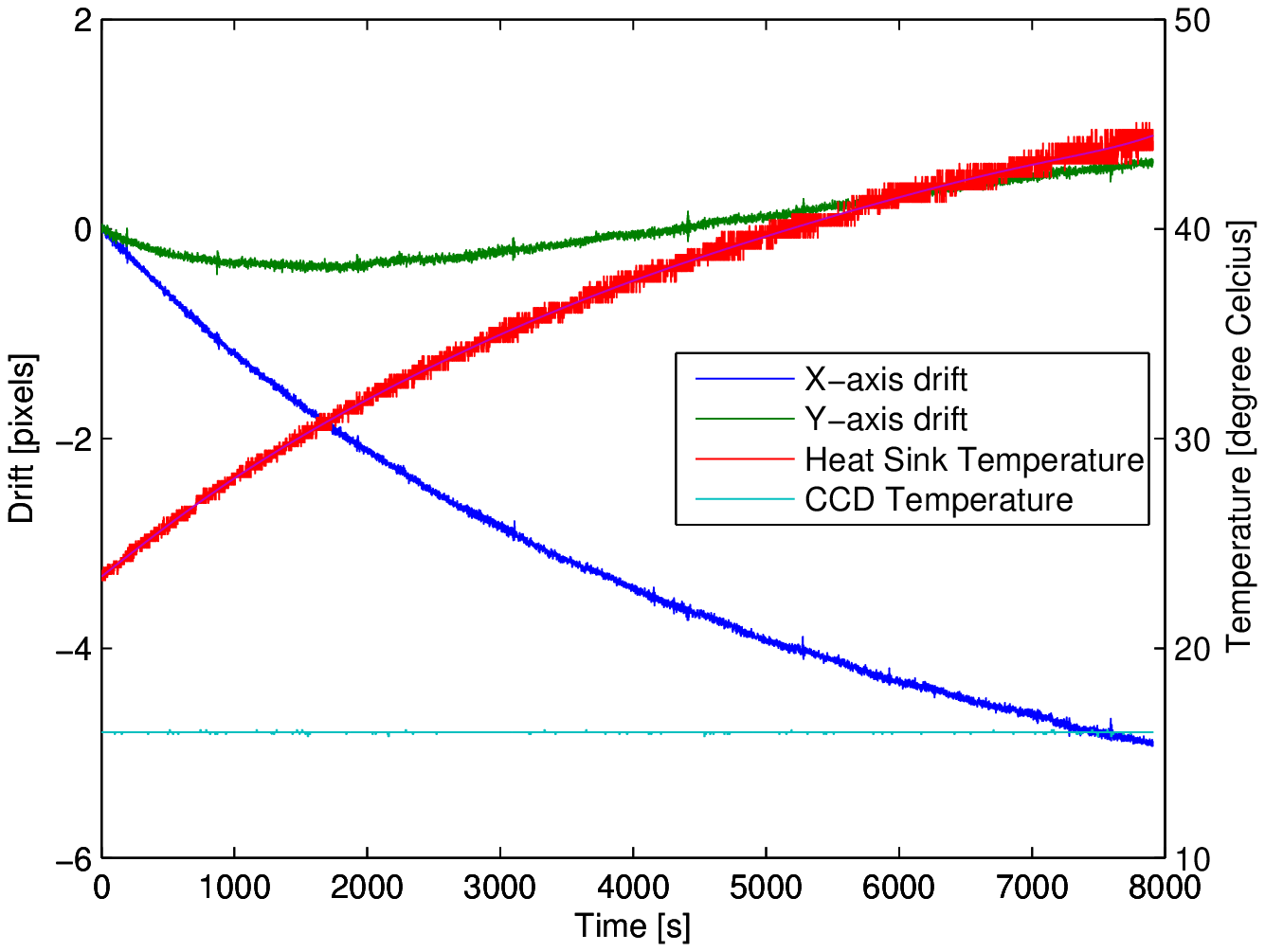}
\caption{Drift and temperature measurements due to overheating of the LOWFS camera (left) and the science camera (right).}%
\label{fig:stdrift}%
\end{figure}

Figure~\ref{fig:stdrift} presents the measured drifts and temperature for the LOWFS camera during one hour (left), and for the science camera during  two hours and ten minutes (right)

In the LOWFS camera, the temperature increased by almost $20^\circ$C in one hour. The maximum acceptable temperature of this camera is $85^\circ$C, so even if it seems that the stable position is below this value, it takes a long time to achieve it, and it is so high that the impact on the noise is not negligible.

In the science camera, a ThermoElectric Cooler (TEC) maintain the detector temperature at a fixed value ($16^\circ$C in that figure), but the heat-sink temperature increases to $45^\circ$C (the maximum value before the camera automatically shuts itself off) in about 2 hours and a half.

In the LOWFS camera, the drift is essentially in y, while in the science camera, the drift is mostly in x. This is coherent with the the way both cameras are mounted on the table: the LOWFS camera is simply on a long vertical stainless steel column, and the science camera is mounted on the side.

The drift measured here is then dominated by the expansion of the camera bodies, as well as the post on which they are mounted. A cooling system was then clearly mandatory to perform wavefront control and validate the required contrast levels during several hours. That is why we implemented a cooling fluid circuit to cool both cameras at the same time.


\section{Long Term Instabilities}
\label{sec:LongTermInstabilities}


\subsection{Drift in the Science Camera}
\label{sec:DriftInTheScienceCamera}

During our latest vacuum test, the cooling system allowed the temperature of both cameras to stay at a stable temperature below the maximum allowed. But even with a stable temperature, we still observe a long term drift.

\begin{figure}[t]%
\center
\FIG{0.5}{false}{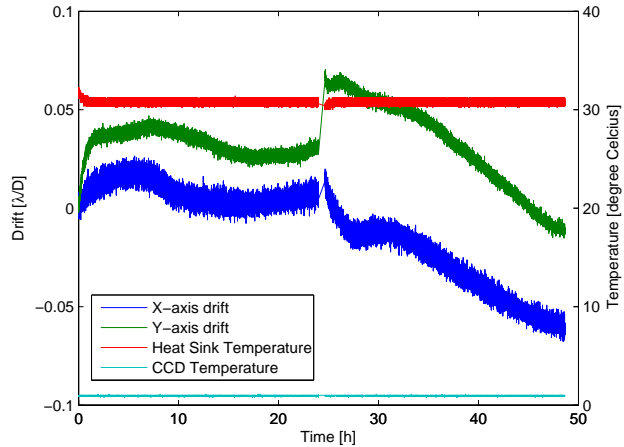}
\caption{Drift of the PSF in the science camera, and temperature measurements of the detector and the heat sink.}%
\label{fig:drift1}%
\end{figure}

Figure~\ref{fig:drift1} presents the drift measured in the science camera, as well as the temperature of the detector and the heat sink, during 48~hours. On this figure, we can see a jump after 24~hours, corresponding to a recalibration of the bench.

While The TEC stabilizes the detector at $1^\circ$C, the liquid cooling system stabilizes the heat sink at $31^\circ$C. The temperature of the heat sink actually varies a little bit when we don't acquire images, leading to a small period of stabilization of the temperature, lasting less than an hour.

Despite a stable temperature of the camera, we can still observe a drift. The first day, the drift was about $4\E{-2}$~\loD peak-to-peak, while the second day, the drift was $1.2\E{-1}$~\loD peak-to-peak. That drift is clearly not due to temperature variations inside the camera, so there is an other disturbance that moves the optical elements or the whole bench. 


\subsection{Low Order Wavefront Sensor and Control with the Deformable Mirror}
\label{sec:LowOrderWavefrontSensorAndControlWithTheDeformableMirror}

The drift described in Sec.~\ref{sec:DriftInTheScienceCamera} was too important for the wavefront control to create the dark hole at the desired contrast level. Therefore, I implemented a slow LOWFSC, using tip/tilt modes of the DM as the actuator. Indeed, unlike the setup in air described in~\cite{Lozi13}, we do not have any piezo stage controlling the source or a fast steering mirror in vacuum, so the only actuator available was the DM.

the LOWFS is set at a frame rate of 100~Hz, and a hundred commands are averaged to send the result to the DM with a frame rate of 1~Hz. With such a low frame rate, we only aim to correct the long-term drift. The gain of the integrator loop is set to 0.1, which is enough for that purpose.

\begin{figure}[b]%
\center
\FIG{0.49}{false}{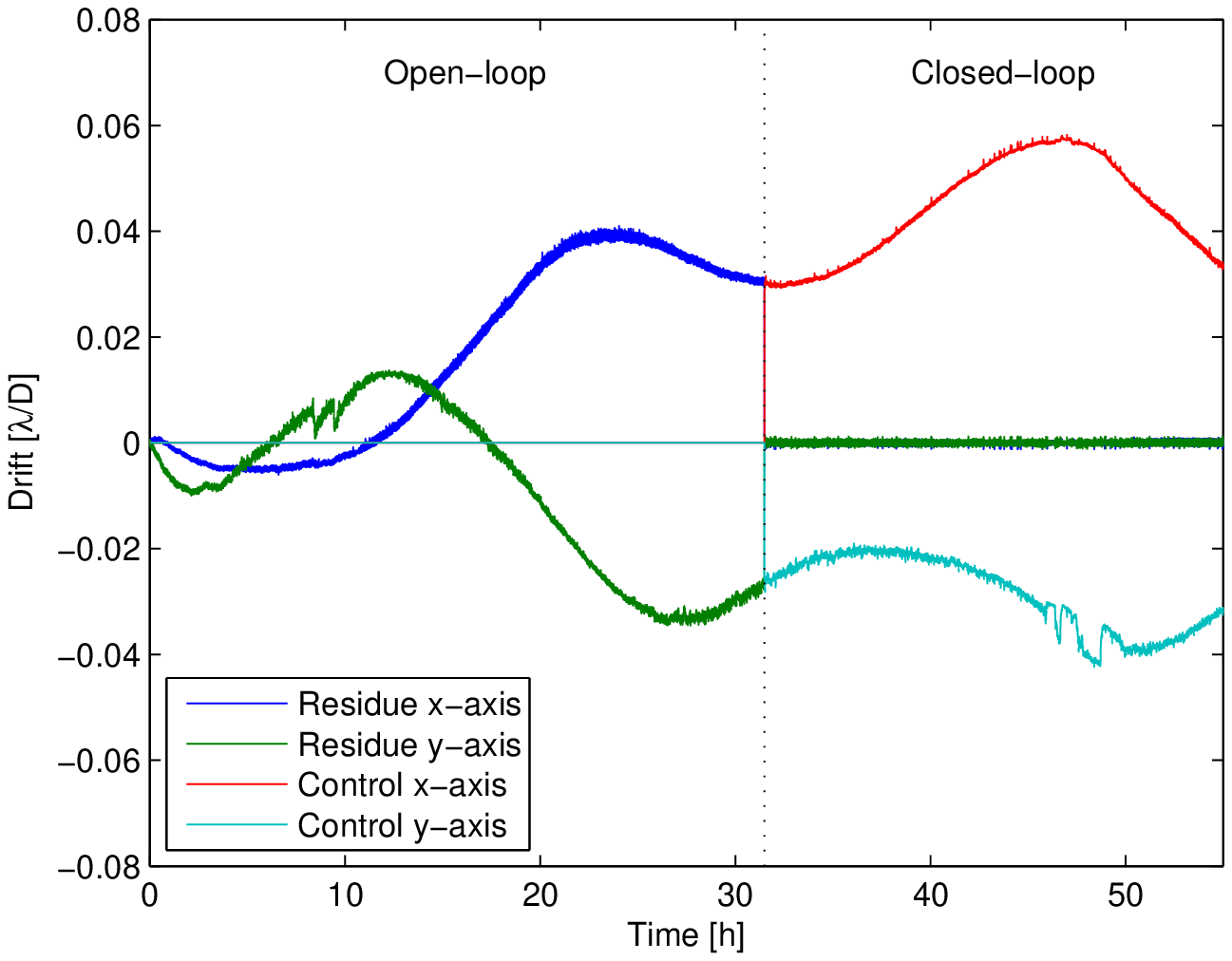}
\FIG{0.49}{false}{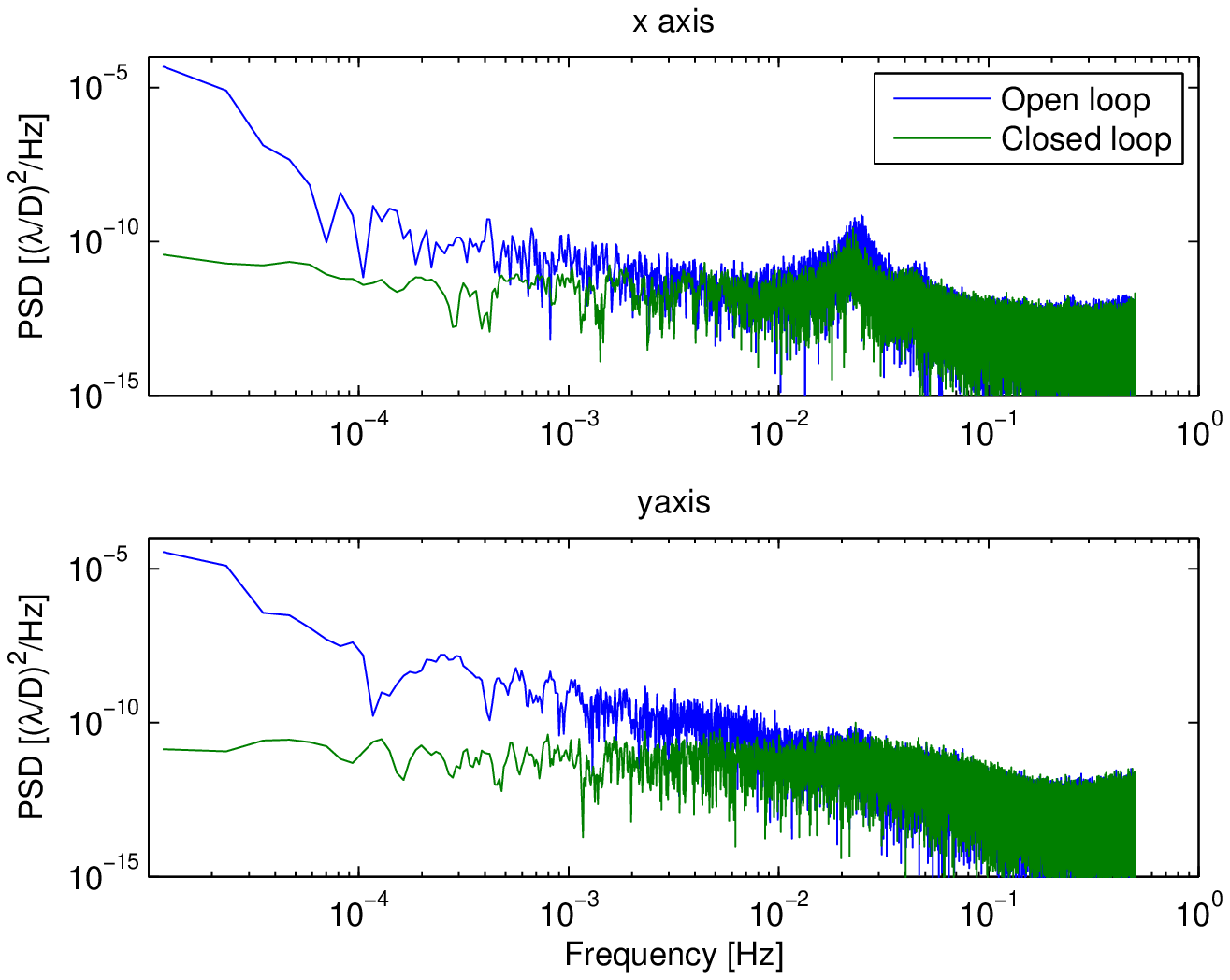}
\caption{Left: Drift measured with the LOWFS in open-loop (for the first 31.5~hours) and in closed loop (between 31.5~h and 55~h), and command applied to the DM. Right: PSDs in open-loop and closed-loop, on both axes.}%
\label{fig:drift2}%
\end{figure}

Figure~\ref{fig:drift2} (left) presents a long measurement of the stability in open-loop between 0 and 31.5~hours, and in closed-loop after 31.5~hours. It also presents the command sent to the DM, converted in units of \loD. We can see that when the loop is closed, the command follows the open-loop drift. The drift has a very low frequency, with what looks like a period of approx. 24~hours. Two hypotheses are privileged to explain this drift: the outside temperature variations make the whole building move, and distort the testbench, or the bench is moving due to the tidal forces created by the moon and the sun. 

On that figure, we also observe jumps, around hour~9 and hour~48, only on the y-axis. It is probably due to mounts releasing stress, or air bubbles popping inside the chamber. But as we can see for the jump around hour~48, they are perfectly managed by the control loop.

On Fig.~\ref{fig:drift2} (right), on the x-axis, both open-loop and closed loop show some kind of damped vibration around a frequency of $2.5\E{-2}$~Hz, i.e. a period of 40~s, and a small harmonics at $5\E{-2}$~Hz. We have not identified what causes that disturbance, but it increase the residual tip by a factor~2.

\begin{figure}[t]%
\center
\FIG{0.49}{false}{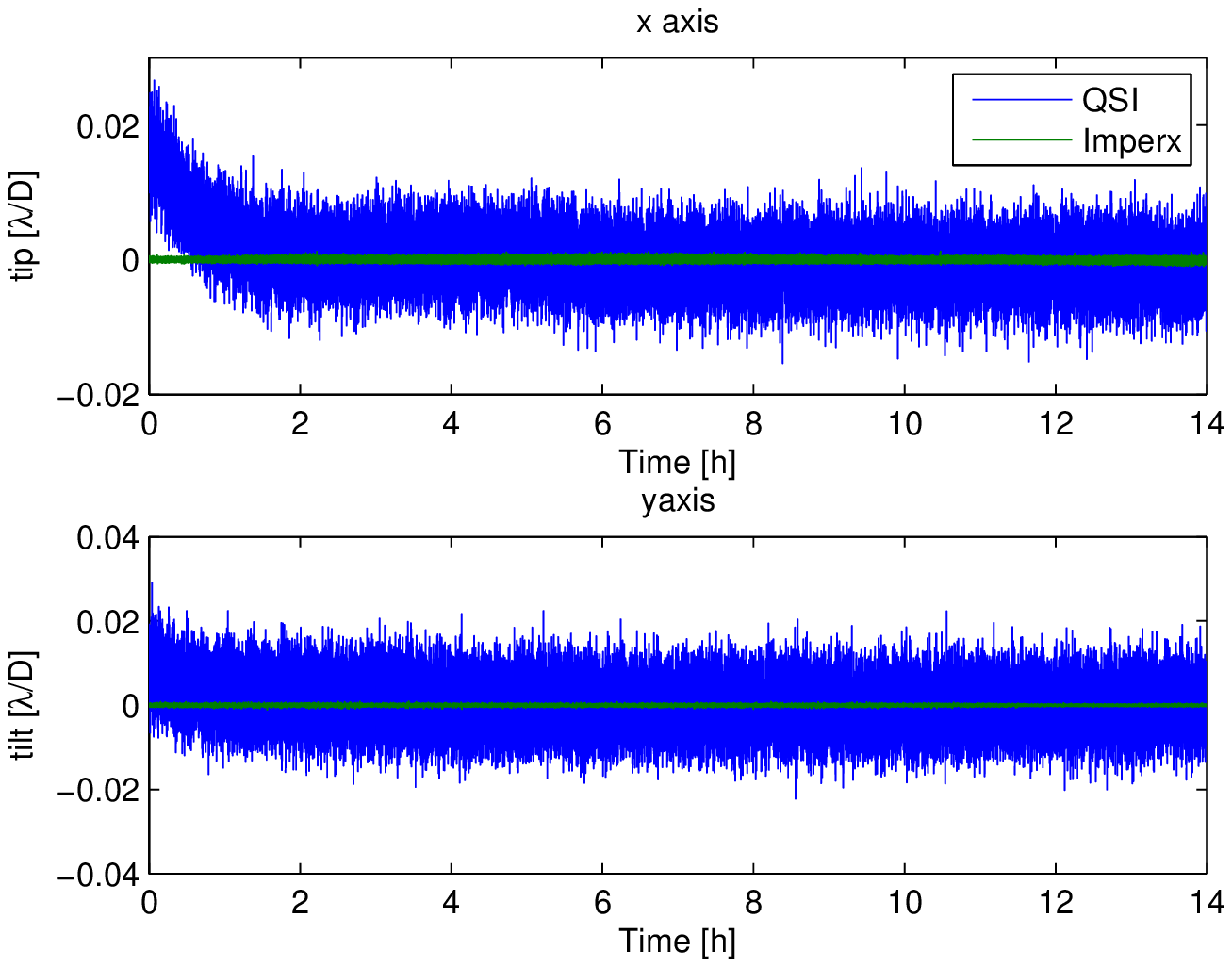}
\FIG{0.49}{false}{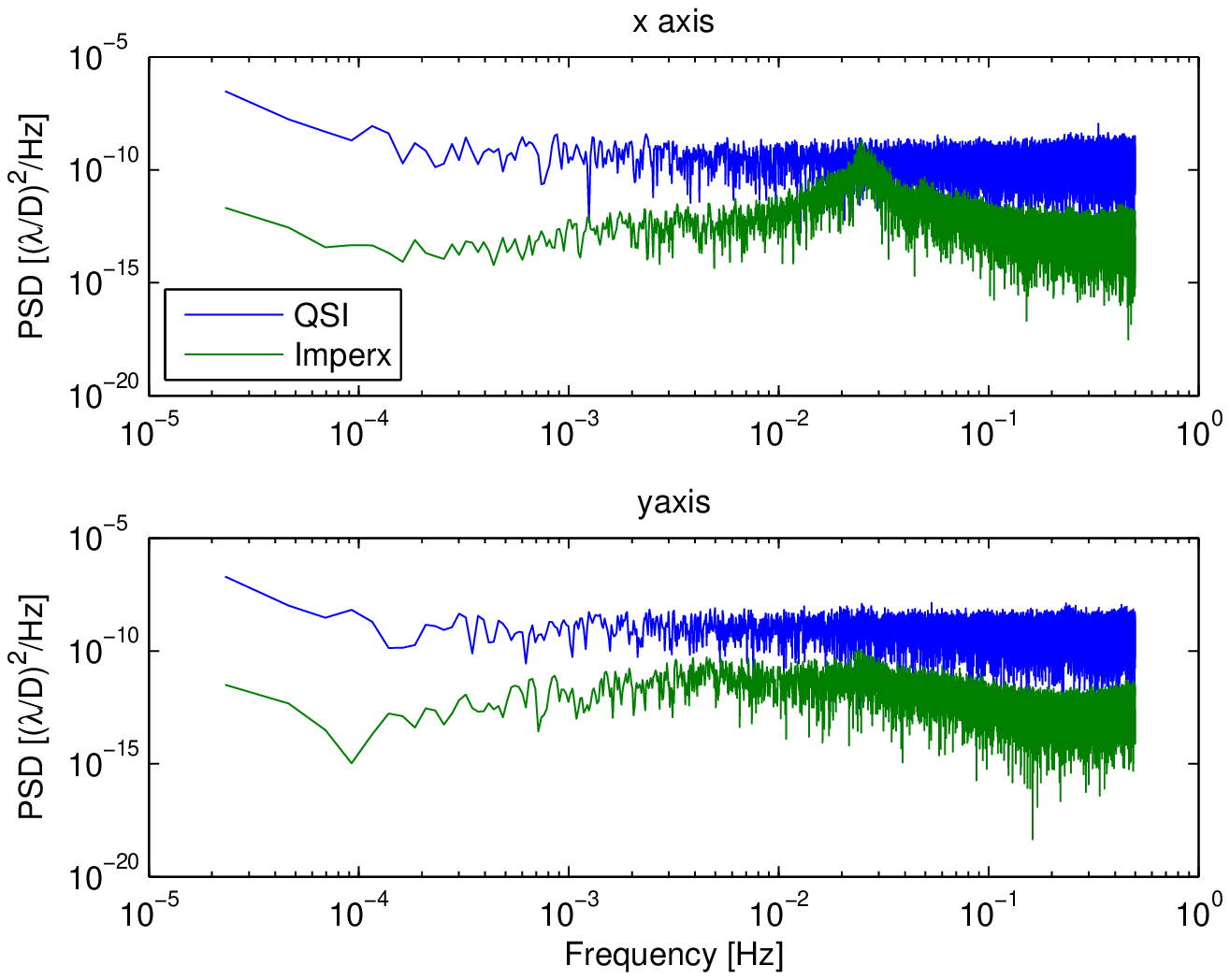}
\caption{Comparison between the residues in closed-loop on the LOWFS camera and the science camera: temporal measurement (left) and PSDs (right).}%
\label{fig:drift3}%
\end{figure}

Figure~\ref{fig:drift3} compares measurements taken with the LOWFS camera and the science camera when the loop is closed, to analyze any non-common path aberrations. First we can observe that the measurements taken with the science are much noisier: its noise level is about $4\E{-3}$~\loD, while it is $\e{-4}$~\loD in the LOWFS. This is due to a different sampling of the PSF in each camera, and the average of 100~images we do in the LOWFS for one applied command. In closed-loop, the residue at 1~Hz is $3.5\E{-4}$~\loD rms in x, and $1.9\E{-4}$~\loD rms in y. Those value are equivalent to the results obtained on the HCIT bench at JPL\cite{Guyon12b}.

In Fig.~\ref{fig:drift3} (left), we can see that the science camera is drifting for about an hour, and becomes very stable after. This drift is typical from the temperature stabilization period of the heat sink, observed in Fig.~\ref{fig:drift1}. But after the camera stabilizes in temperature, no non-common path aberration are seen, at least up to the noise level of the science camera. Of course, this does not guaranty that the FPM is not drifting, but at least we know that the optics after the FPM does not add unseen errors. A three-zone FPM as described in~\cite{Lozi13} should help remove any drift of the mask.


\section{Conclusion}
\label{sec:Conclusion}

This paper presented a successful implementation of a coronagraphic low-order wavefront sensor and control in vacuum. With a simple design, we were able to confirm the results we obtained in air, with a residue  of a few $\e{-3}$~\loD rms in closed-loop, using the DM as the actuator.

Even if the control loop is only running at 1~Hz, having a fast sensor at 1~kHz is very useful to analyze the vibrations that are disrupting the tip/tilt modes. Also, we found out that even in vacuum, there are still very slow drifts that make the LOWFSC essential to perform any kind of wavefront correction on a long time scale.

This LOWFS is very interesting for any type of \coro using a focal plane mask, because its design is very simple, and it can be rapidly implemented on any testbench or instrument. The results we presented here can be scaled to any future mission, for example EXCEDE or AFTA, even if they do not use a PIAA to apodize the beam.

In future developments, we will implement a new three-zone FPM, that will reduce any non-common path aberrations that might limit us. We also want to add a fast steering mirror to perform a much faster correction, up to 1~kHz, and eventually simulate the conditions of space \coros by injecting vibrations and pointing errors.



\bibliography{report}   
\bibliographystyle{spiebib}   

\end{document}